\documentclass[conference]{IEEEtran}
\ifCLASSOPTIONcompsoc
  \usepackage[nocompress]{cite}
\else
  \usepackage{cite}
\fi
\usepackage{cite}
\usepackage{amsmath,amssymb,amsfonts}
\usepackage{algorithmic}
\usepackage{graphicx}
\usepackage{textcomp}
\usepackage{xcolor}
\PassOptionsToPackage{hyphens}{url}\usepackage{hyperref}
\def\BibTeX{{\rm B\kern-.05em{\sc i\kern-.025em b}\kern-.08em
    T\kern-.1667em\lower.7ex\hbox{E}\kern-.125emX}}
\begin{document}

\title{Malware Detection Through Memory Analysis}

\author{\IEEEauthorblockN{Sarah Nassar}
\IEEEauthorblockA{\textit{Electrical and Computer Engineering} \\
\textit{Queen's University}\\
Kingston, Canada}
}

\maketitle

\begin{abstract}
This paper summarizes the research conducted for a malware detection project using the Canadian Institute for Cybersecurity's MalMemAnalysis-2022 dataset. The purpose of the project was to explore the effectiveness and efficiency of machine learning techniques for the task of binary classification (i.e., benign or malicious) as well as multi-class classification to further include three malware sub-types (i.e., benign, ransomware, spyware, or Trojan horse). The XGBoost model type was the final model selected for both tasks due to the trade-off between strong detection capability and fast inference speed. The binary classifier achieved a testing subset accuracy and F1 score of 99.98\%, while the multi-class version reached an accuracy of 87.54\% and an F1 score of 81.26\%, with an average F1 score over the malware sub-types of 75.03\%. In addition to the high modelling performance, XGBoost is also efficient in terms of classification speed. It takes about 37.3 milliseconds to classify 50 samples in sequential order in the binary setting and about 43.2 milliseconds in the multi-class setting. The results from this research project help advance the efforts made towards developing accurate and real-time obfuscated malware detectors for the goal of improving online privacy and safety. \textit{This project was completed as part of ELEC 877 (AI For Cybersecurity) in the Winter 2024 term.}
\end{abstract}


\section{Introduction}\label{section:introduction}

Software systems play an extensive role in most aspects of present-day life. From phones and computers to even financial transactions and classroom learning, these systems serve to simplify, streamline, and augment everyday tasks. However, as with all inventions and technologies, software systems also come with risks. One such risk is malware, which is a general term used to describe any malicious software designed to damage or exploit a programmable device or several devices on a network. Malware can be seen as a kind of infection that can result in different types of harm, including information leakage, service disruption, and financial loss. Therefore, to mitigate the risks associated with malware attacks, accurate real-time detection is a crucial mechanism that warrants research and investment.

\subsection{Motivation}\label{section:motivation}

IBM reported that the global average cost of a data breach in 2023 was \$4.45 million USD \cite{ibm2023cost}. Cybersecurity Ventures estimated in their 2023 almanac that global cybercrime costs will grow by 15\% annually over the next three years, reaching \$10.5 trillion USD annually by 2025 \cite{morgan2023almanac}. According to the World Economic Forum, cyber insecurity now takes the eighth spot in the top 10 rankings of the most severe global risks over the next decade \cite{heading2024risks}. This is the same list that also includes other serious risks like extreme weather events, misinformation and disinformation, adverse outcomes of artificial intelligence (AI) technologies, involuntary migration, and pollution.

In the 2021 results of the Canadian Survey of Cyber Security and Cybercrime, 18\% of participating Canadian businesses reported that they were impacted by cybersecurity incidents \cite{statscan2022impact}. This ratio varied significantly by business size, at 16\% for small businesses (10 to 49 employees), 25\% for medium businesses (50 to 249 employees), and 37\% for large businesses (250 or more employees). 40\% of the impacted businesses experienced downtime as a result, with an average duration of 36 hours, 14\% reported loss of revenue, and 11\% were impacted by ransomware. The total cost of recovery among Canadian businesses impacted by a cybersecurity incident in 2021 summed up to just over \$600 million CAD.

Other than ransomware, which is malware designed to extort money from the targeted system's owner, spyware is another dangerous type of malware that is designed to steal sensitive personal, financial, or organizational data in order to sell them or use them for other malicious purposes. In 2004, America Online (AOL) and the National Cyber Security Alliance (NCSA) conducted the Online Safety Study, which involved asking 329 people some survey questions as well as running scans on their Internet-connected computers \cite{aolncsa2004study}. The results were surprising; although 53\% of respondents said they had spyware or adware, the scans revealed that 80\% had known spyware or adware installed. It is important to note that this study was conducted almost 20 years ago, and with the advancement of malware technologies since then and the increased pervasiveness of the Internet, it is crucial to recognize that spyware is likely to be even more common nowadays. It is also surprising that similar studies were not conducted again more recently.

In 2021, Amnesty International's Security Lab, in collaboration with 80 journalists from 17 media organizations in 10 countries and the University of Toronto research group, Citizen Lab, released a technical report revealing the results of the forensic examination of 67 mobile phones belonging to various people, including human rights activists, journalists, and political dissidents \cite{amnesty2021report}. This investigation found that Pegasus, one of the world's most advanced and invasive spyware tools and the main product of Israeli cyber warfare company, NSO Group, was used to hack 23 phones (34\%) and attempt to hack 14 phones (21\%). For the remaining 30 phones (45\%), the tests were inconclusive, in several cases due to the phones having been replaced. This investigation was initiated by the discovery of a list of 50,000 phone numbers that were potentially selected for targeting by NSO Group's clients \cite{ruiz2021pegasus}. Pegasus is a well-known spyware that can be detected by some antivirus software and caught by digital forensics labs. However, it is a sophisticated malware that can target both iOS and Android mobile devices, even through zero-click attacks, and is continuously being updated to exploit vulnerabilities and infect devices more efficiently while evading detection. In 2019, Facebook launched a lawsuit against NSO Group for exploiting a WhatsApp vulnerability to infect 1,400 user devices with Pegasus \cite{stempel2021facebook}.

Between 2016 and 2018, Citizen Lab scanned the Internet for servers associated to Pegasus, and found 1,091 IP addresses matching the fingerprint and 1,014 domain names pointing to them \cite{marczak2018hideandseek}. Citizen Lab came up with a novel technique to cluster some of the matches to 36 distinct operators. Citizen Lab also designed a global DNS cache probing study on the domain names to identify 45 countries in which Pegasus operators may have been conducting surveillance operations. These include countries with histories of state-sponsored human rights abuses and unethical tactics to silence political dissidents. At least 10 operators appeared to be engaged in cross-border surveillance.

All these examples of the dangerous use of malware and its impacts highlight the necessity of effective detection and prevention tools to make the world a safer place.

\subsection{Problem Formulation}\label{section:problem}

Given the privacy, safety, and financial risks of malware, the goal of this research is to work on the task of accurate and timely detection of obfuscated malware using memory analysis features extracted in the Canadian Institute for Cybersecurity (CIC) MalMemAnalysis-2022 dataset \cite{carrier2022detecting}. Specifically, various machine learning algorithms were compared in two settings: 1) binary classification of benign versus malware software, and 2) multi-class classification to further identify the type of malware from the three options of ransomware, spyware, and Trojan horse, in addition to the benign class.

\subsection{Contributions}\label{section:contributions}

The research described in this paper details three key contributions:

\begin{enumerate}
    \item Improved binary classification performance from the results reported in the original dataset paper at 99.98\% testing subset accuracy and F1 score with an extreme gradient boosting (XGBoost) classifier.
    \item Effective multi-class classification with an accuracy of 87.54\% and an F1 score of 81.26\%, also with an XGBoost classifier.
    \item Extensive classification time analysis showing the efficiency of both the binary classifier (37.3 milliseconds for 50 samples) and multi-class version (43.2 milliseconds for 50 samples).
\end{enumerate}

\section{Related Work}\label{section:relatedwork}

The authors of the dataset used in this project wrote a literature review section in their paper focusing on malware detection using memory analysis \cite{carrier2022detecting}. Although there is a lot of research in the field, there are some significant obstacles, such as advanced malware obfuscation, which is the process of using techniques to make malware remain hidden from detection methods. Manual detection methods or signature-based methods are not very helpful for detecting new unseen malware, so machine learning techniques are promising to achieve a balance between high accuracy, new malware detection capability, and resource consumption efficiency.

Malware detection through static and dynamic methods can be inaccurate. As an alternative, memory analysis is the process of capturing memory snapshots and extracting features from them to analyze the activities running in a system. The dataset's authors proposed some features to extract from the memory dump and achieved a 99.00\% binary classification accuracy and 99.02\% F1 score with a stacking ensemble \cite{carrier2022detecting}. They also approximated that classifying 50 samples in a row with their model takes 400 milliseconds. They compared their method to four other works. The first work introduced a framework for hardware-assisted malware detection based on monitoring and classifying virtual memory access patterns \cite{xu2017malware}. This work had a high detection rate at 99.0\% and medium complexity with a high memory usage and no mention of speed. The second work detected obfuscated malware with a computer vision approach by applying binary-to-image rendering methods to convert memory dumps into RGB images \cite{bozkir2021catch}. This work achieved a high accuracy at 96.39\% and a detection speed of 3.56 seconds per sample. The complexity of the model and memory usage are high because the system needs to store RGB images. The authors of this work also released a dataset called Dumpware10 involving 10 malware classes as well as benign samples for future research. The third work has a model that requires a lower amount of memory with high speed \cite{javaheri2018framework}. However, it does have increased complexity and a lower accuracy at 85\%. The fourth work uses volatile memory dumps and analyzed the data with the MinHash method, which is a technique for quickly estimating how similar two sets are \cite{nissim2019volatile}. The authors used the minimal values of the hash functions as features, and their model achieved up to 100\% true positive rate and as little as 1.8\% for false positive rate. This method has lower complexity compared to the previous two methods, but it is slower and the memory usage is not mentioned. In the dataset paper, the authors highlight that their method achieves a high accuracy, very fast speed, low memory usage, and medium complexity \cite{carrier2022detecting}.

A 2017 survey on malware detection using data mining techniques provides a summary of at least 25 feature extraction methods dating from 2001 to 2016 \cite{ye2017survey}. Most of these methods involved static and/or dynamic analysis and did not focus on memory analysis through memory dumps. The survey also summarized 25 classification papers within the same time range, where the highest accuracy reported was 98.07\% with a support vector machine (SVM) by combining static and dynamic analysis \cite{anderson2012improving}.

A comprehensive review paper published in 2020 outlines six key datasets in the field of malware research released from 2009 to 2018, such as the Microsoft malware classification challenge dataset, the Android Adware and General Malware (AAGM) dataset, and the Elastic Malware Benchmark for Empowering Researchers (EMBER) dataset \cite{aslan2020comprehensive}. This paper notes the prevalence of static and dynamic analysis with no mention of memory analysis. The malware detection approaches were grouped into signature-based, behaviour-based, heuristic-based, model checking-based, deep learning-based, cloud-based, mobile-based, and IoT-based. From a comparison of the malware detection rate of these approaches as the malware complexity increases, the authors note that none of the methods were robust to complex malware, but the behaviour-based approach was in the lead. This underscores the need to develop prevention and detection methods that can keep up with the latest sophisticated malware attacks.

\section{Methodology}\label{section:methodology}

To implement this project, a machine learning pipeline was built, consisting of data exploration, data preparation, class separability visualization, k-fold cross-validation, training with multiple model algorithms, evaluation with key performance metrics, and inference time experiments.

\subsection{Dataset Description}\label{section:datasetdescription}

The dataset used in this research is the CIC MalMemAnalysis-2022 dataset \cite{carrier2022detecting}. The dataset collectors extended the VolMemLyzer (Volatility Memory Analyzer), which is a module that extracts features from memory dumps taken during live malware infection \cite{lashkari2021volmemlyzer}. The extension focuses on obfuscated malware and extracts new features. To test the feature extractor, 2,916 malware samples were collected from VirusTotal and executed in a VirtualBox virtual machine (VM). Benign processes were also executed to simulate normal user behaviour with different applications. Memory dumps were captured from outside the VM on a Windows 10 system using the VirtualBox virtual memory management system every 15 seconds up to 10 times. Over-sampling of the benign records was done with the SMOTE algorithm to make the dataset balanced. Overall, the dataset is composed of 58,596 records, with an even split of 29,298 benign records and 29,298 malicious records.

The CSV dataset file was downloaded from Kaggle at \href{https://www.kaggle.com/datasets/jlcole/cic-malmem-2022}{https://www.kaggle.com/datasets/jlcole/cic-malmem-2022}.

\subsection{Data Preparation}\label{section:datapreparation}

\begin{figure}[t]
\centering
\includegraphics[width=\columnwidth]{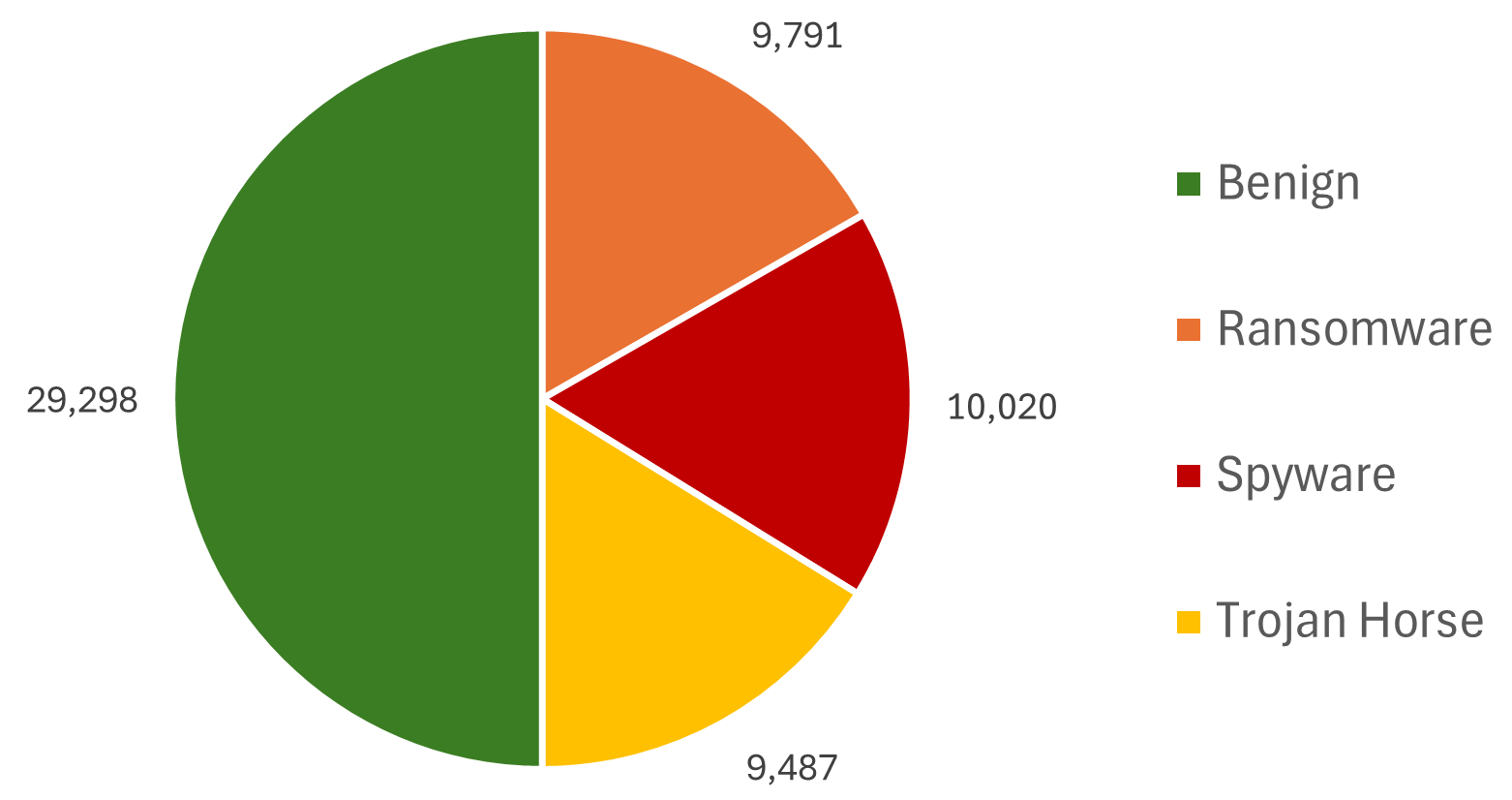}
\caption{Pie chart of record labels.}
\label{fig:piechart}
\end{figure}

Using the Python pandas package to read the file, it was found to have 58,596 rows and 57 columns. One of the columns contained the binary label of benign or malware. As described in the dataset paper, it was confirmed to have 29,298 rows with the benign label and 29,298 rows with the malware label. Another column included the malware category, family, ID, and sample number. The malware category was extracted from this column for the multi-class classification setting. As can be seen in the pie chart in Fig. \ref{fig:piechart}, 10,020 rows were labelled as being spyware, 9,791 rows were labelled as being ransomware, and 9,487 rows were labelled as being Trojan horse malware. The label columns were label encoded to numeric values, as opposed to string values. For the binary case, 'Benign' was encoded as the negative class (0) and 'Malware' was encoded as the positive class (1). For the multi-class case, the labels were encoded according to alphabetical order ('Benign': 0, 'Ransomware': 1, 'Spyware': 2, and 'Trojan': 3).

\begin{figure}[t]
\centering
\includegraphics[width=\columnwidth]{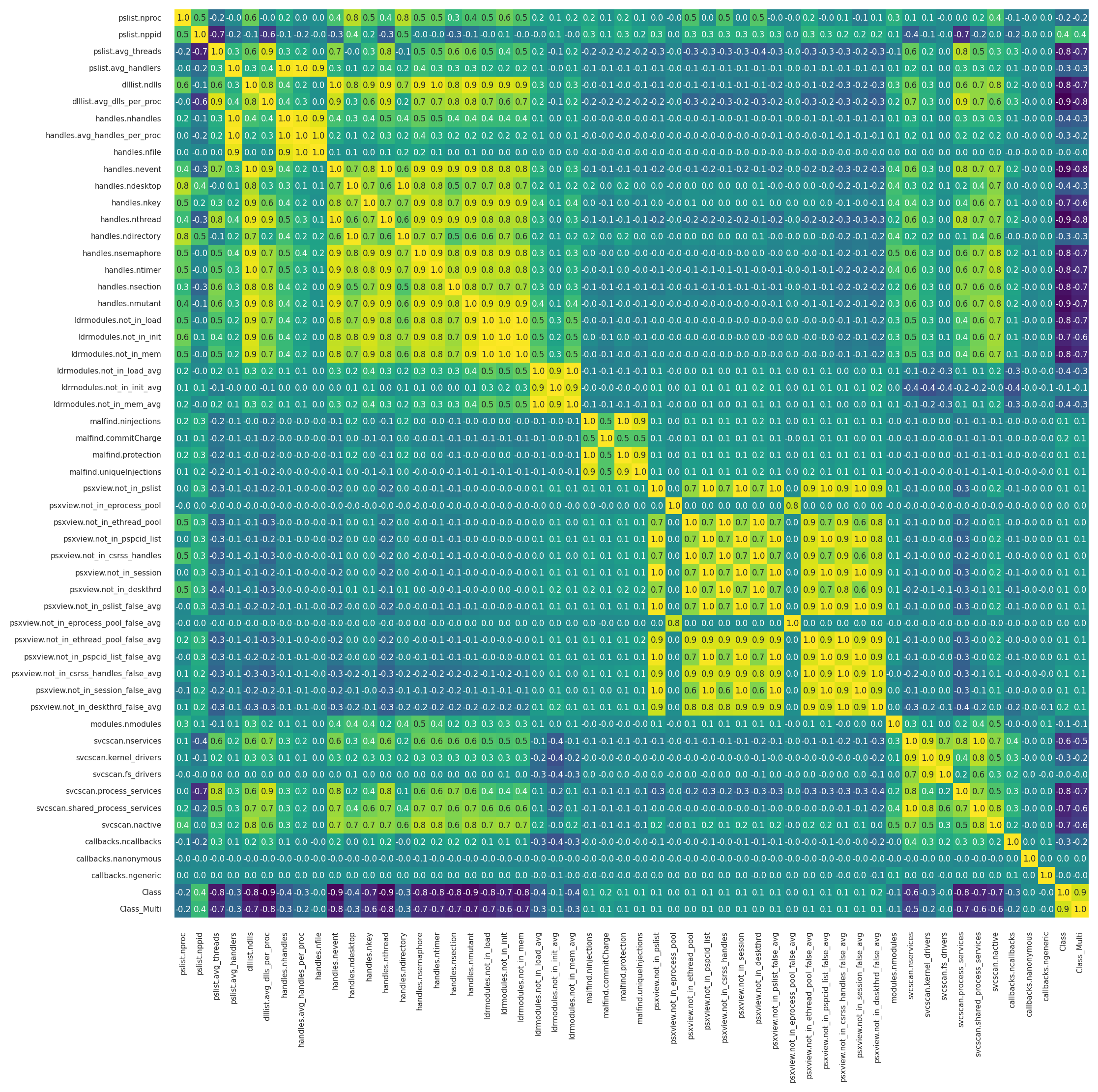}
\caption{Correlation heatmap of the dataset's features. The last two rows and rightmost two columns represent the binary and multi-class labels, respectively.}
\label{fig:heatmap}
\end{figure}

After exploring the pandas DataFrame, it became noticeable that three of the feature columns had a constant value of zero, so they were removed. These columns were pslist.nprocs64bit (total number of 64-bit processes), handles.nport (total number of port handles), and svcscan.interactive\_process\_services (total number of interactive service processes). Excluding the label columns and the constant-valued feature columns, the number of features remaining was 52. To visualize how the features are correlated with both the binary and multi-class labels, as well as how the features are correlated with each other, the pair-wise correlation of columns was calculated with the Pearson correlation coefficient with a pandas function and the resulting matrix was plotted as a heatmap with the seaborn visualization Python library, shown in Fig. \ref{fig:heatmap}. From the correlation heatmap, it is noticeable that less than one-third of the features are highly correlated with the target features and that many of the features have a high correlation with each other.

In order to be able to evaluate the machine learning models, the dataset was split into training and testing subsets with a ratio of 80\% training (46,876 rows) and 20\% testing (11,720 rows). The split was done in a stratified fashion according to the multi-class labels in order to have the same class ratios in both the training and testing subsets.

To make sure the features were all within the same range of values, the min-max scaling method was used to place each feature's values within the range of [0,1]. The scaler was fitted on the training subset, then applied to transform both the training and testing subsets. The procedure was performed in this way to ensure no data leakage in the scaling step.

\begin{figure*}
\centering
\includegraphics[width=\textwidth]{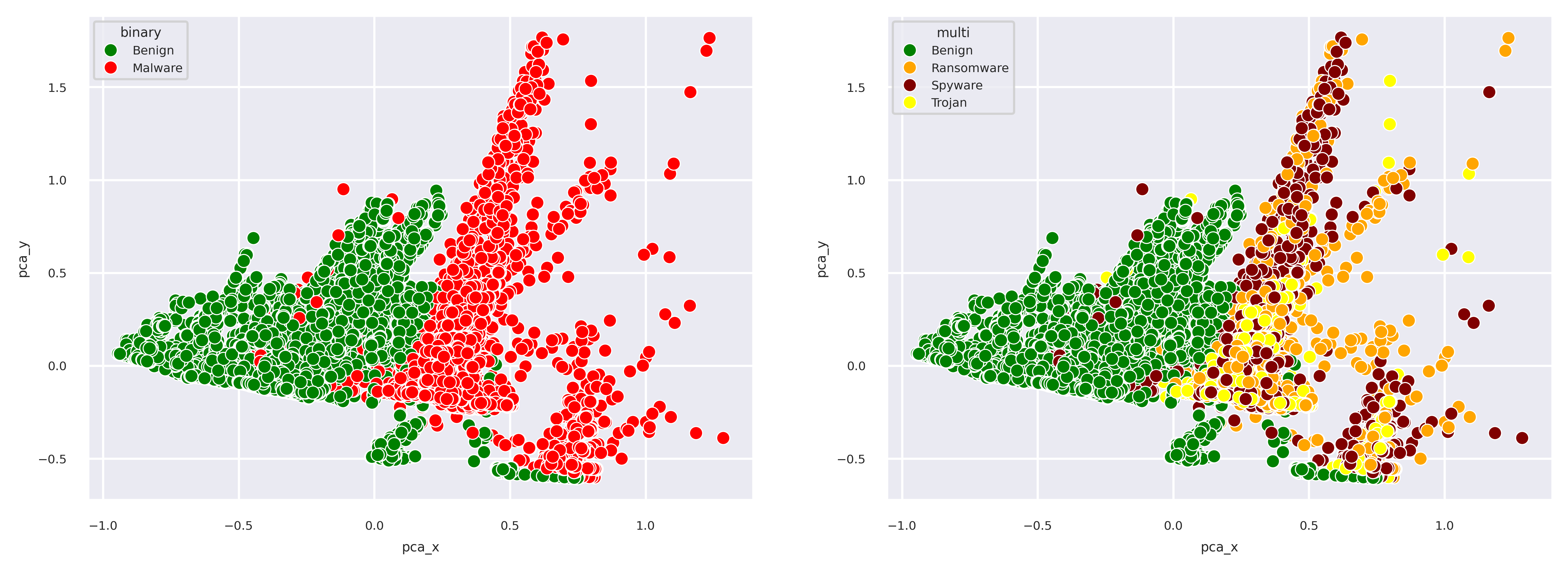}
\caption{Binary (left) and multi-class (right) class separability in 2D through dimensionality reduction with PCA.}
\label{fig:pca}
\end{figure*}

The final step before moving on to modelling was to visualize the class separability in 2D by applying dimensionality reduction to the training features using principal component analysis (PCA) with two components. The components were plotted using seaborn as scatter plots with legends to distinguish class labels. The 2D components are displayed in Fig. \ref{fig:pca} with both the binary and multi-class legends. The benign and malware samples appear to be very distinct as easily separable groups. However, the malware samples among themselves are less distinguishable in terms of separating the ransomware, spyware, and Trojan horse samples.

The dataset split, feature scaling, and dimensionality reduction were all done with the scikit-learn library in Python.

\subsection{Modelling}\label{section:modelling}

The modelling step is made up of two parts in two configurations: for both the binary and multi-class classification configurations, k-fold cross-validation was performed as well as the usual training and testing.

\begin{figure}[t]
\centering
\includegraphics[width=0.65\columnwidth]{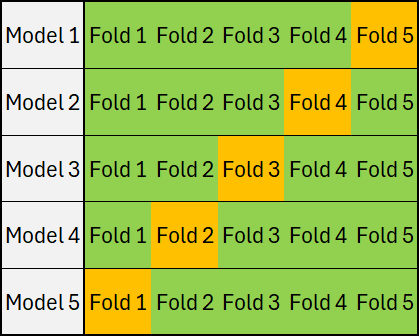}
\caption{Visualization of k-fold cross-validation procedure with k=5 (i.e., five folds). For each model, the four green folds are used for training and the single orange fold is used for evaluation.}
\label{fig:crossval}
\end{figure}

The typical modelling procedure involves training the machine learning model with the training subset, then evaluating the model on the unseen testing subset. The perceived performance from this method can be misleading, because it can potentially be very different if another training and testing split is used (i.e., the random state used for shuffling the data before splitting them). The reason for this is that the model would be seeing different data points during training and then would be evaluated on another set of unseen data points. For example, if there is a dataset with images of cats and birds and the task is to classify them, and some of the images were taken during the day with plenty of light while other were taken during the night in the dark, then the model's ability to generalize would depend on which images were in the training subset and which were in testing subset. If the training subset had only the daytime images, then the model's performance might be very poor on the nighttime images in the testing subset, and vice versa. Therefore, to get a more reliable and complete picture of the model's ability to learn from a dataset, k-fold cross-validation can be used. This method works by splitting the training subset into k partitions, or folds. Then, k models are trained, each with a different combination of k-1 folds, and are each evaluated on an alternating fold that was not used in the training.

k-fold cross-validation was run with five stratified folds. A visualization of the procedure is shown in Fig. \ref{fig:crossval}. To avoid data leakage in the scaling, the original training features were passed (not the scaled ones) and then the min-max scaling procedure was done for each fold separately using the scikit-learn pipeline functionality.

Nine machine learning algorithms were compared. They are described in the following list:

\begin{itemize}
    \item Gaussian naïve Bayes (NB): A probabilistic algorithm based on the Gaussian distribution. Naïve Bayes uses the class frequency (prior), conditional frequencies given the different feature value combinations (likelihood), and feature frequencies (evidence) to calculate the probability that a new data point will have a specific class (posterior). This probability is computed for each class, and the class with the highest probability becomes the model output. This algorithm assumes independence between features to calculate the likelihood as a product of each conditional probability, which is rarely the case, so naïve Bayes is typically only used to check the baseline performance for a given dataset.
    \item Logistic regression (LR): Similar to linear regression, where each input feature is associated with an adjustable coefficient and there is also a learnable intercept term, but used for classification by fitting the learned line to a sigmoid curve to separate the classes. In the binary case, the model output is a number between zero and one. If the output is larger than a threshold, the classification will be the positive class, otherwise it would be the negative class. The default threshold is 0.5, but this can be adjusted.
    \item k-nearest neighbours (KNN): A distance-based algorithm that assigns a new data point to the most frequent class of its k closest neighbours. In this project, k=5 (i.e., the model considers the five nearest neighbours of a new data point in the vote). Various distance metrics can be used, but the most common one is the Euclidean distance.
    \item Multi-layer perceptron (MLP): Another term to refer to feedforward artificial neural network (ANN), which contains artifical neurons in layers that perform a weighted sum of the inputs from the previous layer. These summations then go through an activation function that performs a transformation before being passed to the next layer. The learning mechanism is to compare the network output with the actual output and try to fix the error by performing backpropagation to adjust the weights according to their contribution to the error with partial gradient computations. For binary classification, the final layer's activation should be the sigmoid function to restrict the output to the range of [0,1]. Similarly to the logistic regression decision approach, a threshold is used to decide whether the class should be positive or negative. For multi-class classification, the final layer's activation should be the softmax function to make sure the outputs for the classes add up to one. In this case, the class with the maximum output would become the classification. In this project, there is one hidden layer with 100 artificial neurons, the hidden layer activation is the rectified linear unit (ReLU) function. The model is trained for a maximum of 200 iterations with a batch size of 200 and learning rate of 0.001.
    \item Decision tree (DT): A partitioning algorithm that tries to find the best way to split the data into the separate classes according to their feature values. It has a hierarchical and tree-like structure that can be visualized. For a new data point, it can be traversed down the tree according to its feature values until reaching a leaf node and being classified into the more frequent class.
    \item Random forest (RF): An ensemble of decision trees. It uses the bootstrap aggregating (bagging) ensemble learning method, where each tree is trained using a subset of the dataset, composed through sampling with replacement. After passing a new data point through all the trees, the final output will be based on aggregating each of their outputs. For this project, the random forest is made up of 100 decision trees.
    \item Gradient boosting (GB): An ensemble of decision tree regressors that uses the boosting ensemble learning method. Boosting is a sequential procedure that focuses on improving on samples that were misclassified. Each participating model, called a weak learner, gives higher weight to data points that were incorrectly classified by the previous weak learner to focus on the challenging samples. Gradient boosting uses gradient descent, and weak learners are trained to fit the pseudo-residuals, which indicate the direction to correct the classifications and decrease the error. For this project, 100 weak learners (decision tree regressors) are used.
    \item Light gradient-boosting machine (LGBM): A fast, low-memory consumption, gradient boosting algorithm that uses a histogram-based method to split continuous data into discrete bins and focuses on splitting trees leaf-wise as opposed to level-wise, which may lead to imbalanced trees and overfitting.
    \item Extreme gradient boosting (XGBoost): Similar to gradient boosting, but uses regularization to reduce overfitting and is a faster version.
\end{itemize}

Most of the models were available in the scikit-learn library. However, the LGBM model was imported from the lightgbm library and the XGBoost model was imported from the xgboost library.

\subsection{Evaluation}\label{section:evaluation}

To measure the performance of the classification models, several metrics were calculated, including the accuracy \eqref{equation:accuracy}, balanced accuracy \eqref{equation:balancedaccuracy}, recall \eqref{equation:recall}, precision \eqref{equation:precision}, F1 score \eqref{equation:f1}, Matthews correlation coefficient (MCC), receiver operating characteristic area under the curve (ROCAUC), and logarithmic (log) loss. In the equations below, $TP$ is the number of true positive predictions, $TN$ is the number of true negative predictions, $FP$ is the number of false positive predictions, and $FN$ is the number of false negative predictions.

For the log loss, a smaller number indicates better performance, but for all other metrics, a larger number indicates better performance. For recall, precision, and F1 score, the overall value is based on the macro average of the values from each individual class. The individual class recall, and precision, and F1 scores are also reported for the testing subset evaluation, but not for the k-fold cross-validation. For the multi-class configuration, the ROCAUC is based on the one-vs-rest strategy. For the k-fold cross-validation, the overall metrics are reported as the average of the values from all the models.

\begin{equation}
{Accuracy = \frac{TP + TN}{TP + TN + FP + FN}}
\label{equation:accuracy}
\end{equation}

\begin{equation}
{Balanced Accuracy = 0.5*(\frac{TP}{TP + FN} + \frac{TN}{TN + FP})}
\label{equation:balancedaccuracy}
\end{equation}

\begin{equation}
{Recall = \frac{TP}{TP + FN}}
\label{equation:recall}
\end{equation}

\begin{equation}
{Precision = \frac{TP}{TP + FP}}
\label{equation:precision}
\end{equation}

\begin{equation}
{F1 = \frac{2*TP}{2*TP + FP + FN}}
\label{equation:f1}
\end{equation}

A timing experiment was also done with the final selected binary and multi-class classification models to test the efficiency of real-time malware detection. The experiment involved getting the model classifications of N samples in sequential order, where a different N was used for each run. The N values tested were 10, 50, 100, 150, 200, 250, 300, 500, 1000, and 10000.

Multiple results DataFrames were set up to save the performance metrics and timing results from the experiments for easier comparison later on. These results were rounded to four decimal places and saved to CSV files.

\section{Results, Comparison, and Discussion}\label{section:resultscomparisondiscussion}

The average evaluation metrics from the k-fold cross-validation are reported in Table. \ref{table:crossval}. For binary classification, all models have an average accuracy and F1 score of over 99\%. The average log loss is also very small, with all models except Gaussian naïve Bayes and logistic regression having an average value of less than 0.01. The top performing models are random forest, LGBM, and XGBoost, with an average accuracy, balanced accuracy, recall, precision, and F1 score of 99.99\%, an average MCC of 99.97\%, and an average ROCAUC of 100\%. XGBoost was able to achieve the smallest average log loss at 0.0006. For multi-class classification, it becomes clear that the models perform significantly worse than in the binary case. This was expected since the malware sub-types (ransomware, spyware, and Trojan horse) were not easily separable, as discussed in the 2D PCA dimensionality reduction and visualization commentary. The top performing models are once again random forest, LGBM, and XGBoost by a large margin compared to the other models. Out of these three, the best was XGBoost with an average accuracy of 87.28\%, an average F1 score of 80.85\%, an average ROCAUC of 97\%, and an average log loss of 0.3055.

\begin{table*}
\centering
\caption{Results of k-fold cross-validation.}
\begin{tabular}{|c|c|c|c|c|c|c|c|c|}
\hline
\textbf{Model} & \textbf{Accuracy} & \textbf{Balanced Accuracy} & \textbf{Recall} & \textbf{Precision} & \textbf{F1} & \textbf{MCC} & \textbf{ROCAUC} & \textbf{Log Loss} \\ \hline
\multicolumn{9}{|c|}{\textbf{Binary}} \\ \hline
NB & 0.9928 & 0.9928 & 0.9928 & 0.9928 & 0.9928 & 0.9856 & 0.9961 & 0.2394 \\ \hline
LR & 0.9961 & 0.9961 & 0.9961 & 0.9961 & 0.9961 & 0.9922 & 0.9999 & 0.0140 \\ \hline
KNN & 0.9996 & 0.9996 & 0.9996 & 0.9996 & 0.9996 & 0.9993 & 0.9999 & 0.0058 \\ \hline
MLP & 0.9991 & 0.9991 & 0.9991 & 0.9991 & 0.9991 & 0.9982 & \textbf{1.0000} & 0.0034 \\ \hline
DT & 0.9998 & 0.9998 & 0.9998 & 0.9998 & 0.9998 & 0.9996 & 0.9998 & 0.0069 \\ \hline
RF & \textbf{0.9999} & \textbf{0.9999} & \textbf{0.9999} & \textbf{0.9999} & \textbf{0.9999} & \textbf{0.9997} & \textbf{1.0000} & 0.0009 \\ \hline
GB & 0.9997 & 0.9997 & 0.9997 & 0.9997 & 0.9997 & 0.9994 & 0.9999 & 0.0025 \\ \hline
LGBM & \textbf{0.9999} & \textbf{0.9999} & \textbf{0.9999} & \textbf{0.9999} & \textbf{0.9999} & \textbf{0.9997} & \textbf{1.0000} & 0.0010 \\ \hline
XGBoost & \textbf{0.9999} & \textbf{0.9999} & \textbf{0.9999} & \textbf{0.9999} & \textbf{0.9999} & \textbf{0.9997} & \textbf{1.0000} & \textbf{0.0006} \\ \hline
\multicolumn{9}{|c|}{\textbf{Multi-Class}} \\ \hline
NB & 0.6859 & 0.5403 & 0.5403 & 0.5969 & 0.4705 & 0.5720 & 0.8736 & 7.7018 \\ \hline
LR & 0.7042 & 0.5596 & 0.5596 & 0.5668 & 0.5550 & 0.5606 & 0.8846 & 0.5410 \\ \hline
KNN & 0.7964 & 0.6937 & 0.6937 & 0.6965 & 0.6938 & 0.6949 & 0.9119 & 1.9233 \\ \hline
MLP & 0.7671 & 0.6523 & 0.6523 & 0.6612 & 0.6481 & 0.6541 & 0.9199 & 0.4699 \\ \hline
DT & 0.8413 & 0.7619 & 0.7619 & 0.7620 & 0.7619 & 0.7620 & 0.8572 & 5.7170 \\ \hline
RF & 0.8698 & 0.8044 & 0.8044 & 0.8044 & 0.8044 & 0.8047 & 0.9672 & 0.3657 \\ \hline
GB & 0.8236 & 0.7355 & 0.7355 & 0.7369 & 0.7355 & 0.7357 & 0.9485 & 0.4013 \\ \hline
LGBM & 0.8678 & 0.8013 & 0.8013 & 0.8012 & 0.8012 & 0.8018 & 0.9678 & 0.3198 \\ \hline
XGBoost & \textbf{0.8728} & \textbf{0.8087} & \textbf{0.8087} & \textbf{0.8085} & \textbf{0.8085} & \textbf{0.8092} & \textbf{0.9700} & \textbf{0.3055} \\ \hline
\end{tabular}
\label{table:crossval}
\end{table*}

The evaluation metrics from the testing subset evaluation are reported in Table \ref{table:testing}. For binary classification, all models perform very well with mostly very close metric scores. The best models are random forest, gradient boosting, and LGBM, with an accuracy, balanced accuracy, recall, precision, F1 score, MCC, and ROCAUC of 100\%, an MCC of 99.97\%, and a ROCAUC of 100\%. LGBM was able to achieve the smallest log loss at zero. For multi-class classification, the top performing models are random forest, LGBM, and XGBoost, significantly improving over the other models. Out of these three, the best was random forest with an accuracy of 87.84\% and an F1 score of 81.72\%. XGBoost was able to achieve a better ROCAUC at 97.09\% and a log loss of 0.3009.

\begin{table*}
\centering
\caption{Results of testing subset evaluation.}
\begin{tabular}{|c|c|c|c|c|c|c|c|c|}
\hline
\textbf{Model} & \textbf{Accuracy} & \textbf{Balanced Accuracy} & \textbf{Recall} & \textbf{Precision} & \textbf{F1} & \textbf{MCC} & \textbf{ROCAUC} & \textbf{Log Loss} \\ \hline
\multicolumn{9}{|c|}{\textbf{Binary}} \\ \hline
NB & 0.9925 & 0.9925 & 0.9925 & 0.9925 & 0.9925 & 0.9850 & 0.9958 & 0.2456 \\ \hline
LR & 0.9966 & 0.9966 & 0.9966 & 0.9966 & 0.9966 & 0.9932 & 0.9998 & 0.0155 \\ \hline
KNN & 0.9994 & 0.9994 & 0.9994 & 0.9994 & 0.9994 & 0.9988 & 0.9999 & 0.0038 \\ \hline
MLP & 0.9994 & 0.9994 & 0.9994 & 0.9994 & 0.9994 & 0.9988 & 0.9998 & 0.0052 \\ \hline
DT & 0.9999 & 0.9999 & 0.9999 & 0.9999 & 0.9999 & 0.9998 & 0.9999 & 0.0031 \\ \hline
RF & \textbf{1.0000} & \textbf{1.0000} & \textbf{1.0000} & \textbf{1.0000} & \textbf{1.0000} & \textbf{1.0000} & \textbf{1.0000} & 0.0006 \\ \hline
GB & \textbf{1.0000} & \textbf{1.0000} & \textbf{1.0000} & \textbf{1.0000} & \textbf{1.0000} & \textbf{1.0000} & \textbf{1.0000} & 0.0003 \\ \hline
LGBM & \textbf{1.0000} & \textbf{1.0000} & \textbf{1.0000} & \textbf{1.0000} & \textbf{1.0000} & \textbf{1.0000} & \textbf{1.0000} & \textbf{0.0000} \\ \hline
XGBoost & 0.9998 & 0.9998 & 0.9998 & 0.9998 & 0.9998 & 0.9997 & \textbf{1.0000} & 0.0003 \\ \hline
\multicolumn{9}{|c|}{\textbf{Multi-Class}} \\ \hline
NB & 0.6814 & 0.5340 & 0.5340 & 0.5713 & 0.4651 & 0.5649 & 0.8738 & 7.9320 \\ \hline
LR & 0.7060 & 0.5626 & 0.5626 & 0.5734 & 0.5527 & 0.5662 & 0.8850 & 0.5431 \\ \hline
KNN & 0.8017 & 0.7020 & 0.7020 & 0.7040 & 0.7022 & 0.7027 & 0.9157 & 1.8602 \\ \hline
MLP & 0.7677 & 0.6533 & 0.6533 & 0.6637 & 0.6396 & 0.6592 & 0.9195 & 0.4711 \\ \hline
DT & 0.8504 & 0.7754 & 0.7754 & 0.7759 & 0.7756 & 0.7756 & 0.8653 & 5.3882 \\ \hline
RF & \textbf{0.8784} & \textbf{0.8172} & \textbf{0.8172} & \textbf{0.8172} & \textbf{0.8172} & \textbf{0.8176} & 0.9697 & 0.3390 \\ \hline
GB & 0.8218 & 0.7328 & 0.7328 & 0.7342 & 0.7328 & 0.7331 & 0.9483 & 0.4021 \\ \hline
LGBM & 0.8676 & 0.8009 & 0.8009 & 0.8008 & 0.8009 & 0.8014 & 0.9683 & 0.3174 \\ \hline
XGBoost & 0.8754 & 0.8127 & 0.8127 & 0.8127 & 0.8126 & 0.8131 & \textbf{0.9709} & \textbf{0.3009} \\ \hline
\end{tabular}
\label{table:testing}
\end{table*}

From the multi-class classification testing evaluation, the per-class recall, precision, and F1 score for the three malware classes are shown in Table \ref{table:perclass}. The per-class metrics are important to look at because the multi-class classification task has imbalanced data (i.e., many more benign samples), so the high performance on the benign samples dominates the overall metrics and makes them appear overly good. Random forest was able to achieve the largest values in all cases except the spyware recall. The ransomware recall was 73.54\% and the Trojan horse recall was 73.76\%. XGBoost had the best spyware recall at 80.19\%. In terms of precision, random forest achieved a ransomware precision of 73.51\%, a spyware precision of 79.20\%, and a Trojan horse precision of 74.19\%. Random forest also reached a ransomware F1 score of 73.53\%, a spyware F1 score of 79.39\%, and a Trojan horse F1 score of 73.98\%.

\begin{table*}
\centering
\caption{Per-class recall, precision, and F1 score for the malware classes from the testing subset evaluation (R: ransomware, S: spyware, T: Trojan horse).}
\begin{tabular}{|c|c|c|c|c|c|c|c|c|c|}
\hline
\textbf{Model} & \textbf{Recall\_R} & \textbf{Recall\_S} & \textbf{Recall\_T} & \textbf{Precision\_R} & \textbf{Precision\_S} & \textbf{Precision\_T} & \textbf{F1\_R} & \textbf{F1\_S} & \textbf{F1\_T} \\ \hline
NB & 0.0358 & 0.2206 & 0.8936 & 0.4895 & 0.4433 & 0.3538 & 0.0666 & 0.2946 & 0.5069 \\ \hline
LR & 0.2610 & 0.3792 & 0.6138 & 0.4776 & 0.4388 & 0.3811 & 0.3375 & 0.4069 & 0.4702 \\ \hline
KNN & 0.6103 & 0.6452 & 0.5527 & 0.5605 & 0.6354 & 0.6211 & 0.5844 & 0.6403 & 0.5849 \\ \hline
MLP & 0.2952 & 0.5634 & 0.7550 & 0.5809 & 0.5874 & 0.4869 & 0.3915 & 0.5751 & 0.5920 \\ \hline
DT & 0.6956 & 0.7255 & 0.6807 & 0.6726 & 0.7388 & 0.6928 & 0.6839 & 0.7321 & 0.6867 \\ \hline
RF & \textbf{0.7354} & 0.7959 & \textbf{0.7376} & \textbf{0.7351} & \textbf{0.7920} & \textbf{0.7419} & \textbf{0.7353} & \textbf{0.7939} & \textbf{0.7398} \\ \hline
GB & 0.5904 & 0.6742 & 0.6670 & 0.6218 & 0.7171 & 0.5983 & 0.6057 & 0.6950 & 0.6308 \\ \hline
LGBM & 0.7114 & 0.7789 & 0.7134 & 0.7203 & 0.7648 & 0.7183 & 0.7158 & 0.7718 & 0.7158 \\ \hline
XGBoost & 0.7257 & \textbf{0.8019} & 0.7234 & 0.7347 & 0.7771 & 0.7394 & 0.7302 & 0.7893 & 0.7313 \\ \hline
\end{tabular}
\label{table:perclass}
\end{table*}

To balance between model performance and time efficiency, and based on the high k-fold cross-validation performance, the final model selected was XGBoost for both binary and multi-class classification. The dataset compilers reported in their paper that their best performance for binary classification was an accuracy of 99.00\% and an F1 score of 99.02\%. Their model was a stacking ensemble where the base learner were naïve Bayes, random forest, and decision tree and the meta learner was logistic regression. In this work, an improved performance was achieved, where the XGBoost classifier has an accuracy and F1 score of 99.98\%. The dataset collectors did not report any experiment results for multi-class classification, so this work appears to be one of the first exploring this task, where the XGBoost accuracy is 87.54\% and the F1 score is 81.26\%. The macro average of the recall and F1 score of the malware sub-types only (excluding the benign class) are 75.03\%, while for precision, the macro average is 75.04\%.

\begin{figure}[t]
\centering
\includegraphics[width=0.65\columnwidth]{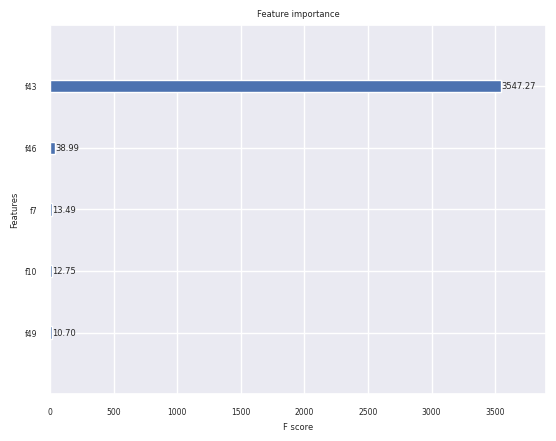}
\caption{Top five important features for the XGBoost binary classifier.}
\label{fig:featureimportancebinary}
\end{figure}

\begin{figure}[t]
\centering
\includegraphics[width=0.65\columnwidth]{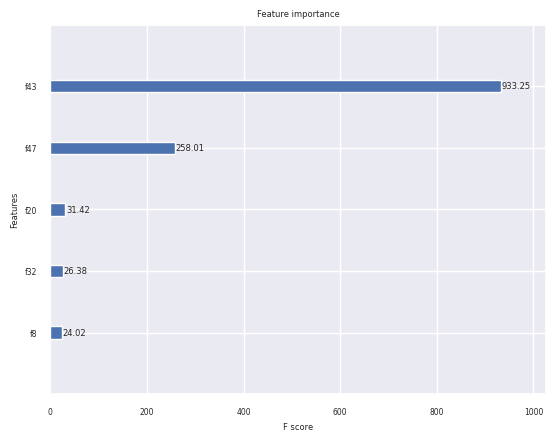}
\caption{Top five important features for the XGBoost multi-class classifier.}
\label{fig:featureimportancemulticlass}
\end{figure}

The feature importance, measured as the average gain of splits which use each feature, was calculated. The top five important features for binary classification are shown in Fig. \ref{fig:featureimportancebinary} and for multi-class classification are shown in Fig. \ref{fig:featureimportancemulticlass}. In both cases, the most important feature appears to be the one at index 43, which is svcscan.nservices (total number of services). In the binary case, the rest of the features appear to be significantly less important. In the multi-class case, the next most important feature seems to be the one at index 47, which is svcscan.shared\_process\_services (total number of Windows 32 shared processes). From the correlation heatmap commentary, it was noted that not all features have a high correlation with the target labels and that several features have a high correlation with each other. Some preliminary experiments with feature selection and feature reduction showed a drop in performance, but only slightly.

The timing experiment results can be seen in Table \ref{table:time}. For each number of samples tested in the results reported by the dataset collectors, the binary classification times in this work are much faster. Although the dataset compilers did not run experiment with multi-class classification models, the mutli-class XGBoost classifier in this work was still very fast and only a little slower than the binary classifier. Classifying 50 samples in a row took 400 milliseconds for the dataset authors, but took only 37.3 milliseconds in the binary case here and 43.2 milliseconds in the multi-class case.

Although the binary model performance and classification speed in this work improve over the results reported in the dataset's paper, it is important to note that the authors of that paper did not mention their training and testing subset split for reproducibility and comparison. Also, a difference in computer specifications would play a part in the difference in classification speed. In this work, a Python Google Compute Backend Engine in Google Colaboratory was used that had 12.7GB of available RAM. One more thing to note is that the timing analysis only includes the classification speed and not the amount of time required to save a memory dump, extract the features from it, and prepare the record before passing it to the model (e.g., min-max scaling).

\begin{table}
\centering
\caption{Results of the timing experiments with XGBoost in seconds.}
\begin{tabular}{|c|c|c|}
\hline
\textbf{Number of Samples} & \textbf{Binary Model} & \textbf{Multi-Class Model} \\ \hline
10 & 0.0064 & 0.0098 \\ \hline
50 & 0.0373 & 0.0432 \\ \hline
100 & 0.0678 & 0.0844 \\ \hline
150 & 0.3429 & 0.8769 \\ \hline
200 & 0.1693 & 0.2265 \\ \hline
250 & 0.0856 & 0.1226 \\ \hline
300 & 0.0979 & 0.1151 \\ \hline
500 & 0.1724 & 0.2120 \\ \hline
1000 & 0.3465 & 0.3988 \\ \hline
10000 & 3.4867 & 6.2382 \\ \hline
\end{tabular}
\label{table:time}
\end{table}

\section{Limitations and Future Work}\label{section:limitationsfuturework}

The main limitations of this work are related to the dataset. The complete dataset was not available on the CIC's official website, so a version was found on Kaggle. The CSV file's set of features did not appear to be fully in line with those described in the paper. For example, the API-hook features were not found (total number of apihooks, total number of in line apihooks, and total number of apihooks in user mode). Further, the choice of malware sub-types is a little unusual, specifically the choice to include Trojan horse malware alongside ransomware and spyware. While ransomware and spyware are malware goals or types, Trojan horse is a malware implementation style that involves being disguised as a legitimate program. This means that the Trojan horse malware style can be used to implement ransomware or spyware. So, it would have made more sense for the malware sub-types to be more distinct (i.e., either the malware goals or the malware implementation styles).

Another limitation of this work is the problem of data leakage between the training and testing subsets. The dataset compilers collected multiple memory dumps from the same samples separated by 15 seconds up to 10 times. This resulted in 29,298 malware records from 2,916 malware samples and 29,298 benign records from an undisclosed number of benign applications. To avoid data leakage, it would be better to separate the records such that all records from a single sample are fully in either the training or testing subset, but not a mix of both because this could result in learning some characteristics inherent to a specific sample then testing on the same sample again, which can make the testing subset performance appear better and result in invalid conclusions about model generalizability to other samples. A future work can be to use the record numbers to ensure each sample's records are only placed in one of the subsets. With the current dataset format, this would be possible for the malicious samples, but not the benign ones, because the benign sample do not have a unique identifier associated with them and a number for each record.

In terms of future work, the primary focus can be on investigating the feature selection and feature reduction steps to reduce the time required to process a memory dump even more. Although the classification speed is already fast, reducing the feature dimensionality can make it even faster and can also reduce the time required for feature extraction. Another possible future direction is to explore the detection capability on the malware families, not only the sub-types. The dataset includes five families within each malware sub-type, such as Conti for ransomware and 180Solutions for spyware. Accurately detecting the malware family can help further improve the defense and recovery capacity.

For deployment, there can be more analysis on the resource consumption, especially when running the detector in real-time, including the time, memory, and storage consumption. For example, the memory dump file is 2GB, which is large. An end-to-end detection pipeline can also be developed that first notifies the user if a program is malware with high accuracy, then show the probabilities or model confidences of the estimated sub-types because the performance is not as high for that task.

\section{Conclusion}\label{section:conclusion}

Overall, the purpose of this research was to explore the effectiveness of machine learning techniques in performing the task of malware detection using the CIC MalMemAnalysis-2022 dataset. Beyond binary classification, multi-class classification models were also investigated for the task of classifying three malware sub-types (ransomware, spyware, and Trojan horse). For both cases, the final model type selected was XGBoost for its high modelling performance and fast classification time. The binary classification version achieved a testing subset accuracy and F1 score of 99.98\%, while the multi-class version reached an accuracy of 87.54\% and an F1 score of 81.26\%. Excluding the benign class performance, the multi-class model achieved an average F1 score of the malware sub-types of 75.03\%. In addition to the effective model performance in distinguishing between benign and malware samples and estimating the malware type, XGBoost is also efficient in terms of classification speed. For example, to classify 50 samples in sequential order, the binary classifier takes about 37.3 milliseconds and the multi-class classifier takes about 43.2 milliseconds. This research helps further the efforts towards the development of accurate real-time malware detectors to enhance online privacy and safety.

\bibliographystyle{IEEEtran}
\bibliography{references}

\end{document}